\documentclass[reprint, aps, showpacs]{revtex4-1}
\usepackage[utf8]{inputenc}
\usepackage[T1]{fontenc}
\usepackage[english]{babel}
\usepackage{amsmath}
\usepackage{amsfonts}
\usepackage{amssymb}
\usepackage{graphicx}

\newcommand{\D}{\mathrm{d}}
\newcommand{\del}{\partial}

\newcommand{\mini}{\mathrm{min}}
\newcommand{\maxi}{\mathrm{max}}
\newcommand{\eq}{\mathrm{eq}}
\newcommand{\crit}{\mathrm{cr}}

\begin{document}
	\title{Positron acceleration via laser-augmented blowouts in two-column plasma structures}
	\author{Lars Reichwein}
	\email{lars.reichwein@hhu.de}
	\affiliation{Institut f\"{u}r Theoretische Physik I, Heinrich-Heine-Universit\"{a}t D\"{u}sseldorf, 40225 D\"{u}sseldorf, Germany}
	
	\author{Anton Golovanov}
	\affiliation{Institute of Applied Physics RAS, 603950 Nizhny Novgorod, Russia}
	\author{Igor Yu. Kostyukov}
	\affiliation{Institute of Applied Physics RAS, 603950 Nizhny Novgorod, Russia}
	
	\author{Alexander Pukhov}
	\affiliation{Institut f\"{u}r Theoretische Physik I, Heinrich-Heine-Universit\"{a}t D\"{u}sseldorf, 40225 D\"{u}sseldorf, Germany}

	\date{\today}
	\begin{abstract}
		We propose a setup for positron acceleration consisting of an electron driver and a laser pulse creating a two-fold plasma column structure. The resulting wakefield is capable of accelerating positron bunches over long distances even when evolution of the driver is considered. The scheme is studied by means of particle-in-cell simulations. Further, the analytical expression for the accelerating and focusing fields are obtained, showing the equilibrium lines along which the witness bunch is accelerated.
	\end{abstract}
	\maketitle
	
	\section{Introduction}
	The efficient acceleration of positrons is of great interest for fundamental physics, especially for the electron-positron colliders \cite{Doebert2018}.
	In the field of accelerator physics, laser-plasma based schemes have specifically grown in interest over the last few decades, as the higher achievable field strengths compared to conventional accelerators allow for higher energies over shorter acceleration distances.

	Acceleration in plasma wakefield can be accomplished either with a driving laser pulse \cite{Tajima1979, Pukhov2002} or a driving particle bunch \cite{Chen1985}. 
	For the laser-driven setup, a so-called bubble can be excited for a normalized laser vector potential $a_0 > 1$, while for the electron-bunch driven setup a sufficiently narrow and high-density bunch has to be used. In both cases, the driver pushes out electrons in the direction transverse to its propagation leaving behind an electronic cavity (``bubble''). Electrons can be injected into the back part of the bubble via several mechanisms, such as self-injection \cite{Mangles2012}, ionization injection \cite{Zeng2014}, or density down-ramp injection \cite{Ekerfelt2017}. For appropriately chosen laser-plasma parameters, the excited bubble exhibits uniform accelerating fields that propagate with a velocity close to the speed of light $c$, which allows for the acceleration of mono-energetic electron bunches \cite{Faure2004}. Utilizing more complicated setups like Trojan horse injection \cite{Hidding2012}, emittances in the order of nm rad and energy spreads in the 0.1\% range were achieved. Currently realizable in the case of homogeneous plasma are electron bunches of up to 8 GeV \cite{Gonsalves2019}.
	
	In the case of positrons, wakefield acceleration proves to be a more difficult challenge: the accelerating and focusing region for positrons in non-linear wakefields is comparatively small. Thus, more advanced setups are necessary. One of the proposed setups utilizes donut-shaped electron bunches as drivers \cite{Jain2015}. Other setups use drive bunches consisting of positrons or hollow plasma targets \cite{Corde2015, Gessner2016, Silva2021}.
	While these schemes improve on the extent of the accelerating region, some still pose problems either in experimental feasibility or stability.
	
	In a recent work \cite{Diederichs2019}, Diederichs \textit{et al.} have proposed a setup using a laser-ionized plasma column of finite radius that is penetrated by an electron driver. Such columns have been shown to be experimentally realizable for radii in the range of several tens of micrometers and peak densities of ca. $10^{17}$ cm$^{-3}$ \cite{Green2014, Shalloo2019}. Due to the column's finite radius, the field structure is different from the case of wakefields in homogeneous plasma. As observed there, the finite plasma column benefits the acceleration of Gaussian positron bunches. What remains critical, however, is the evolution of the electron driver which commonly leads to negative effects when accelerating positrons. A follow-up study of beam loading in this regime which incorporated an algorithm for optimizing the bunch profile was presented in \cite{Diederichs2020}.
	Still, in experiments the effects of beam evolution are likely to negatively influence the efficient acceleration over long distances.
	
\begin{figure}
		\includegraphics[width=\columnwidth]{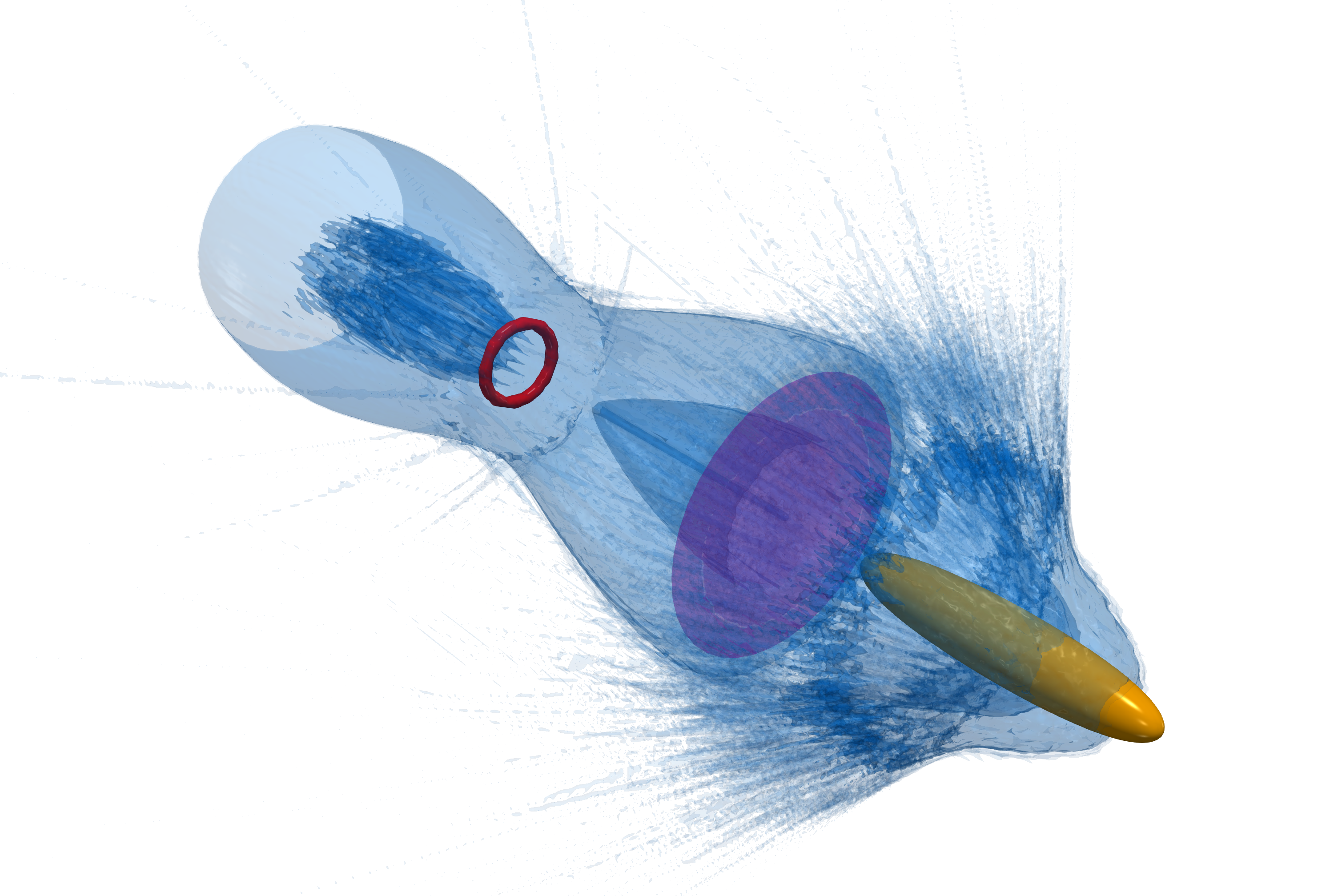}
		\caption{\label{fig:3d}Three-dimensional rendering of the LAB scheme. The electron driver (orange) pushes the electrons (blue) out in the transverse direction and the second column is ionized by the laser pulse (magenta). The positron ring (red) is located in the vicinity of the region with the highest electron density.}
	\end{figure}

	In this paper, we present a novel laser-augmented blowout (LAB) scheme for the acceleration of positrons over distances of tens of centimeters. An electron drive bunch ionizes a narrow column of hydrogen gas and excites a wakefield. A separate laser pulse ionizes a wider column behind the driver. This creates a region beneficial for the acceleration of positron rings (Fig. \ref{fig:3d}). 
	This ring bunch geometry further has the advantage that it will avoid beamstrahlung similarly to flat beams proposed for high-energy colliders like CLIC and is still suitable for acceleration in plasma much like round bunches \cite{Tomassini2019}.
	We will first present our scheme in terms of particle-in-cell (PIC) simulations. Further, the accelerating and focusing fields are derived analytically. The field structure exhibits equilibrium lines along which the positron ring is located. Finally, the emittance and the energy spectrum of the witness bunch are studied. Matching the positron ring according to the setup parameters yields a reduction of the emittance growth.
	
	\begin{figure*}[t]
		\includegraphics[width=\textwidth]{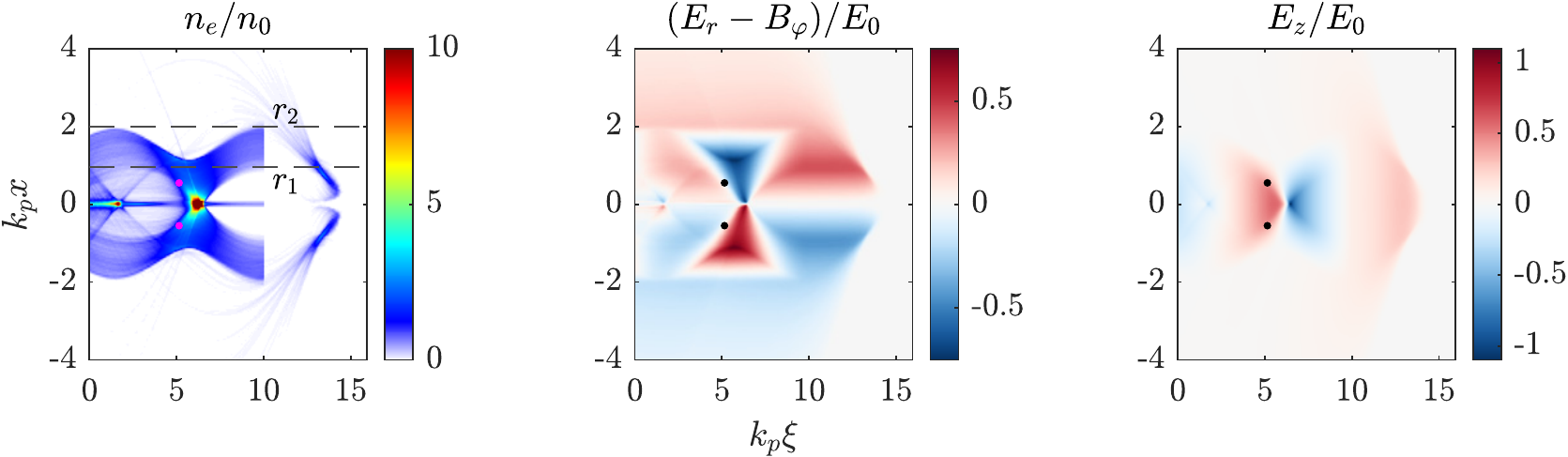}
		\caption{\label{fig:setup}PIC simulation results for the LAB scheme with an exemplary $w_0 = 17$ $\mu$m laser pulse (the later setup uses a focal spot size of $w_0 = 125$ $\mu$m). (a) The electron density (clipped at $n = 10 n_0$ for better visibility) shows the fork-like structure. The values $r_1$ and $r_2$ denote the radii used in the analytical derivation of the fields. As seen in (b), this location of the positrons coincides with the equilibrium line for the focusing force as well as the area where the accelerating gradient of $E_z$ is the largest (c). The magenta/black dots denote the position of the positron ring.}
	\end{figure*}
	
	\section{Setup}
	For our simulations we use the quasistatic PIC code \textsc{qv3d} \cite{Pukhov2016}. 
	In our presented example (Fig. \ref{fig:setup}), our simulation box has dimensions $16 \times 20 \times 20 k_p^{-3}$ with grid size $h_x = h_y = h_z = 0.05 k_p^{-1}$ where $k_p = 2\pi / \lambda_p$ is the plasma wave number ($\lambda_p = 75 \times 10^{-4}$ cm).
	
	The box is filled with initially unionized hydrogen gas with a density of $n_0 = 2 \times 10^{17}$ cm$^{-3}$. A Gaussian-shaped electron driver with dimensions $\sigma_z = 12 $ $\mu$m ($z$ being the direction of propagation) and $\sigma_x = \sigma_y = 2.4$ $\mu$m performs the initial ionization. It has a peak density of $20n_0$ and the initial momentum of the electrons is $p_z = 10^4 m_e c$ with 5\% longitudinal spread ($m_e$ being the electron rest mass). The driver excites a wakefield whose fields exhibit a different structure compared to the case of homogeneous plasma (we will see this in the analytic theory later on in more detail).
	
	A short ($\tau_0 = 4.5$ fs) circularly polarized laser pulse with wavelength $\lambda_L = 400$ nm ionizes a second, wider column. We consider acceleration up to $2\times 10^4 T_0$. As the laser evolves a lot over this scale, we choose a focal spot size of $w_0 = 125$ $\mu$m which gives us a Rayleigh length in the cm range.
    The normalized laser vector potential is $a_0 = 0.025$ throughout the simulations which is sufficient for the ionization of the second column.
	Such laser pulses could be created at the LWS-20 laser which has 16 TW peak power and 70-75 mJ pulse energy \cite{Rivas2017}. Different radii may also be used: Presented in Fig. \ref{fig:setup} we exemplarily show the setup for a pulse with $w_0 = 17$ $\mu$m for better visibility of the parameters used in the analytic discussion. Such a small focal spot would, however, come at the cost of a much shorter Rayleigh length. Therefore, we will stay with $w_0 = 125$ $\mu$m throughout the following simulations. The pulse length may also be increased, but the front of the second column becomes increasingly curved due to this. Choosing a shorter laser pulse also comes with the indirect advantage that, due to the OPCPA frontends typically used by such setups, unwanted ionization by a pre-pulse is reduced. The laser pulse is described using the envelope model; the description of tunnel ionization is according to \cite{Andreev2001,Massimo2020}. Refraction of the laser pulse over time is incorporated in the simulations; further phenomena that could influence the stability of the scheme are discussed later.
	
	We observe that, in the region right behind the electron driver, fields reminiscent of typical blowout structures in homogeneous plasma can be seen. Since only a narrow part of the total gas in the simulation box is ionized, the fields extend over a greater distance leading to a slightly modified blowout structure. Specifically in the region from $\xi = 5 k_p^{-1}$ to $\xi = 7 k_p^{-1}$, the diagonal white lines in the structure of the focusing force $E_r - B_\varphi$ prove to be beneficial for positron acceleration. Further, a look at the accelerating field $E_z$ shows that along these lines, the positrons will experience nearly a maximum accelerating field from the subsequent blowout structure.
	
	Thus, we place the positron ring accordingly (cf. Fig. \ref{fig:3d}). Its density is varied in the range of $1 n_0 $ to $80 n_0$ for the different performed simulations. Its initial radius is 6 $\mu$m with a thickness of 1 $\mu$m and the positrons have an initial momentum of $p_z = 10^4 m_e c$. Such positron rings could be created in a similar fashion to hollow electron bunches via the utilization of Laguerre-Gaussian laser pulses \cite{Baumann2018, Zhang2016}.
	
	As we will now also analytically verify, this location corresponds to an equilibrium line where the radial electric field $E_r$ vanishes and, at the same time, the highest accelerating field for that transverse position is reached. A strongly mismatched positron beam would lead to the loss of positrons and worse emittance.

    \section{Analytical model}
	To obtain the structure of the electromagnetic fields of the LAB scheme analytically, we proceed in a similar fashion to the approach in \cite{Golovanov2017} using the following plasma setup in the quasi-static approximation
	\begin{align}
		&\rho_e (0, r) = \begin{cases}
			-1 \; , & r_1 < r < r_2 \\
			0 \; , & \text{ else}
		\end{cases} \; , \\
		&\rho_i (\xi, r) = \begin{cases}
		1 \; , & r < r_2 \\
		0 \; , & \text{ else}
	\end{cases} \; ,
	\end{align}
	where $\xi = t - z$ denotes the co-moving coordinate and $r_1, r_2$ indicate the column radii (cmp. Fig.  \ref{fig:setup}).
	The position $\xi = 0$ corresponds to the point where the laser pulse ionizes the plasma ring $r_1 < r < r_2$, and we neglect the contribution of electrons ionized and expelled by the electron driver (assuming they are expelled to $r > r_2$).
	Ions are considered immobile, but plasma electrons begin to converge to the axis $r$ due to the action of the radial electric field $E_r$ from the ion column, which can be calculated as
	\begin{equation}
	    E_r(\xi, r) = \frac{1}{r} \int_0^r [\rho_i(\xi, r') + \rho_e(\xi, r')] r' \D r' \; .
	\end{equation}
	Here, we assume that the motion of the electrons is predominantly radial and neglect the influence of the longitudinal field $E_z$ on them, so they do not generate longitudinal current $j_z$ and azimuthal magnetic field $B_\varphi$.
	We also assume that the motion is non-relativistic and that electron trajectories never cross, so the motion of the electron with the initial coordinate $r_0$ is always affected only by the Coulomb field of the electrons with initial coordinates smaller than $r_0$, and we get
    \begin{equation}
        \frac{\D^2 r}{\D \xi^2} = -E_r = - \frac{r}{2} + \frac{r_0^2 - r_1^2}{2r} \;, 
    \end{equation}
    where the first term is the Coulomb attraction from the ions, and the second term is the Coulomb repulsion from the inner electrons.
    If we further assume that $|r - r_0| \ll r_0$ and $r_0 \gg r_1$, we obtain
    \begin{align}
		r = r_0 - \frac{r_1^2}{2 r_0} (1 - \cos \xi) \; .
		\label{eq:trajectory}
	\end{align}
	All electrons oscillate with the same period equal to the plasma wavelength in this solution.
	As the oscillation amplitude decreases with $r_0$, trajectory crossing never happens for this solution.
	Let us now assume that this solution is valid for all electrons with initial coordinates in the interval $[r_1, r_2]$, i.e. for a given $\xi$ all electrons are located between
	\begin{equation}
		r_\mini = \frac{r_1}{2} (1 + \cos \xi) \; , \quad r_\maxi = r_2 - \frac{r_1^2}{2 r_2} (1 - \cos \xi) \; . \label{eq:minmax}
	\end{equation}
	Knowing all trajectories allows us to calculate the density distribution and thus the transverse electric field
	\begin{align}
		E_r (\xi, r) = \begin{cases}
			\frac{r}{2} \; , & r < r_\mini (\xi) \\
			\frac{r}{2} - \frac{r_0^2(\xi, r) - r_1^2}{2r} \; ,&  r_\mini < r < r_\maxi \\
			\frac{r}{2} - \frac{r_2^2 - r_1^2}{2r} \; , & r_\maxi (\xi) < r < r_2 \\
			\frac{r_1^2}{2r} \; , & r > r_2 \; .
		\end{cases} \label{eq:Er}
	\end{align}
	Here we introduced
	\begin{equation}
		r_0 (\xi, r) = \frac{r + \sqrt{r^2 + 2r_1^2 (1 - \cos \xi)}}{2} \; ,
	\end{equation}
	which is the electron's initial coordinate $r_0$ expressed through its current radial coordinate $r$.
	For $r  < r_\mini$ we observe the electric field typical for long ion channels \cite{Whittum1990, Rosenzweig1991}. Outside of the plasma column width, $r > r_2$, we see that $E_r \propto r^{-1}$, which will be valid until electrons expelled from $r < r_1$ begin to contribute to $E_r$.
	
	In turn, the accelerating field is calculated from the Panofsky--Wenzel theorem \cite{Panofsky1956}, $\del E_z/\del r = \del E_r/\del \xi$,
	\begin{widetext}
		\begin{align}
			E_z = \begin{cases}
				E_{z,0}(\xi) - \frac{r_1^2 \sin \xi}{4} \ln \left[ \frac{r_2 r_\maxi}{r_1 r_\mini} \right] \; , & r < r_\mini (\xi) \\
				E_{z,0}(\xi) - \frac{r_1^2 \sin \xi}{4} \ln \left[ \frac{r_2 r_\maxi}{r_0(\xi, r) r} \right] \; , & r_\mini (\xi) < r < r_\maxi (\xi) \\
				E_{z,0}(\xi) \; , & r > r_\maxi (\xi) \; ,
			\end{cases}
			\label{eq:E_z}
		\end{align}
	\end{widetext}
	where $E_{z,0}$ is the additional longitudinal field created by the electrons expelled by the driver.
	For simplicity, we assume that $E_{z,0} = 0$.
	These fields now allow us to calculate the aforementioned equilibrium lines along which the positron ring will be located.  First, we find the line where $E_r = 0$. Using equation (\ref{eq:Er}), we obtain
	\begin{align}
		r_\eq = r_1 \frac{1 + \cos \xi}{2 \sqrt{- \cos \xi}} \; . \label{eq:req}
	\end{align}
	If we linearize $E_r$ along this line, we will see that it is focussing for positrons, thus making it possible to confine positrons in the transverse direction along this line.
	By putting our solution for $r_\eq$ into the expression for $E_z$, we get
	\begin{equation}
		E_{z,\eq} = - \frac{r_1^2 \sin \xi}{4} \ln \left[ \frac{-4 \cos \xi}{1 - \cos^2 \xi} \frac{r_2^2}{r_1^2} - \frac{2 \cos \xi}{1 + \cos \xi} \right] \; . \label{eq:ezeq}
	\end{equation}
    This distribution is symmetrical around $\xi = \pi$.
    As we can see from the equations above, the equilbrium line itself does not depend on $r_2$, but $E_{z,\eq}$ does. If we want to achieve the highest possible accelerating gradient, it is therefore beneficial to increase $r_2$. However, the scaling $E_{z,\eq} \propto \ln (r_2 / r_1)$ is comparatively weak, and it is valid only until $r_2$ does not reach the radius at which the electrons expelled by the driver are.

\begin{figure}
    \includegraphics[width=\columnwidth]{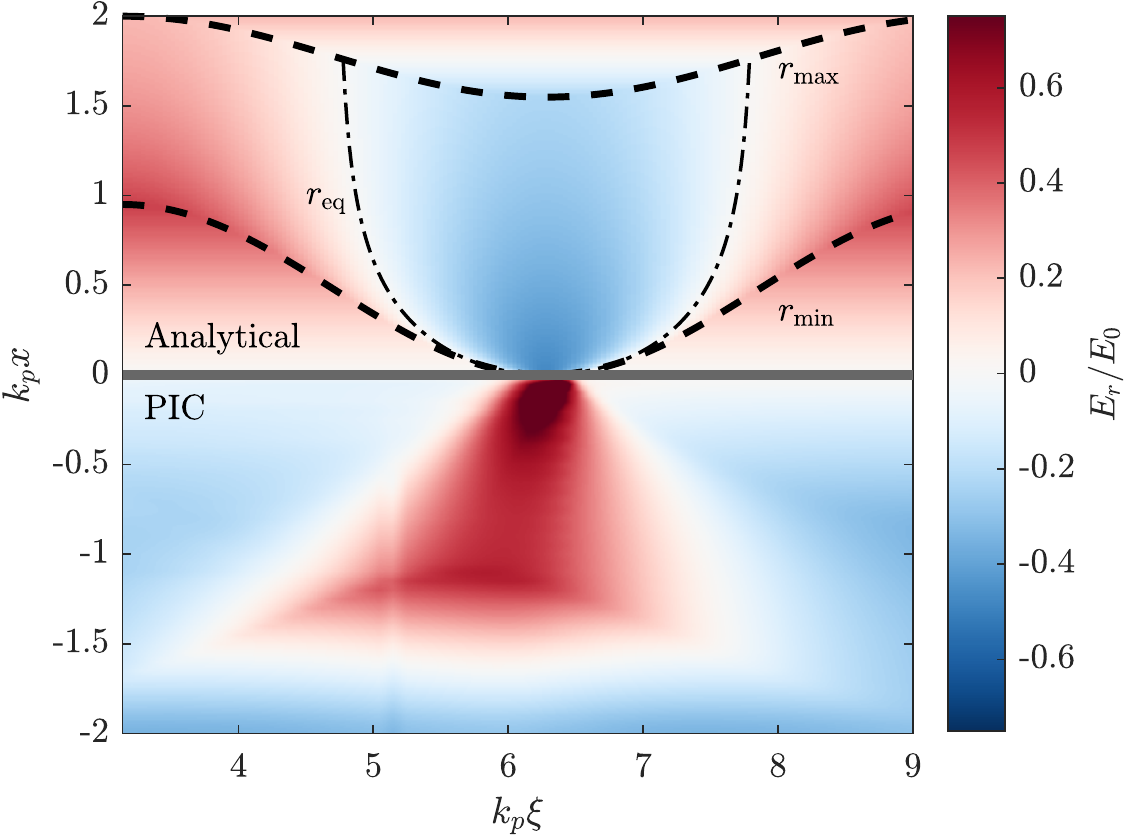}
    \caption{\label{fig:cmp}Comparison of the analytical expression and the simulations results for the structure of $E_r$ for the smaller focal spot size laser from Fig. \ref{fig:setup}. The dashed lines denote the lines $r_\mathrm{min}$ and $r_\mathrm{max}$ from eq. (\ref{eq:minmax}) that parameterize the borders of the different field regions and the dashed-dotted line denotes the equilibrium line $r_\mathrm{eq}$ along which the positron ring can be accelerated from eq. (\ref{eq:req}).}
\end{figure}

\section{Simulation results}
A comparison of our analytical result with the simulation can be seen in Fig. \ref{fig:cmp}. The analytical structure qualitatively agrees very well with the numerical result: Along the dashed-dotted equilibrium line $r_\mathrm{eq}(\xi)$ the positron ring can be accelerated.
According to Eq.~\eqref{eq:E_z}, the value of $E_z$ is constant between $0$ and $r_\mini$, but then begins to decrease with $r$.
However, as $r_\eq$ stays fairly close to $r_\mini$, the acceleration of positrons is achieved in almost maximum possible field.
The field structure in the simulations is not symmetric, however, because of the limit of applicability of the assumptions we made, as the real oscillation period of electrons depends on their initial radial position.

\begin{figure}
	\includegraphics[width=\columnwidth]{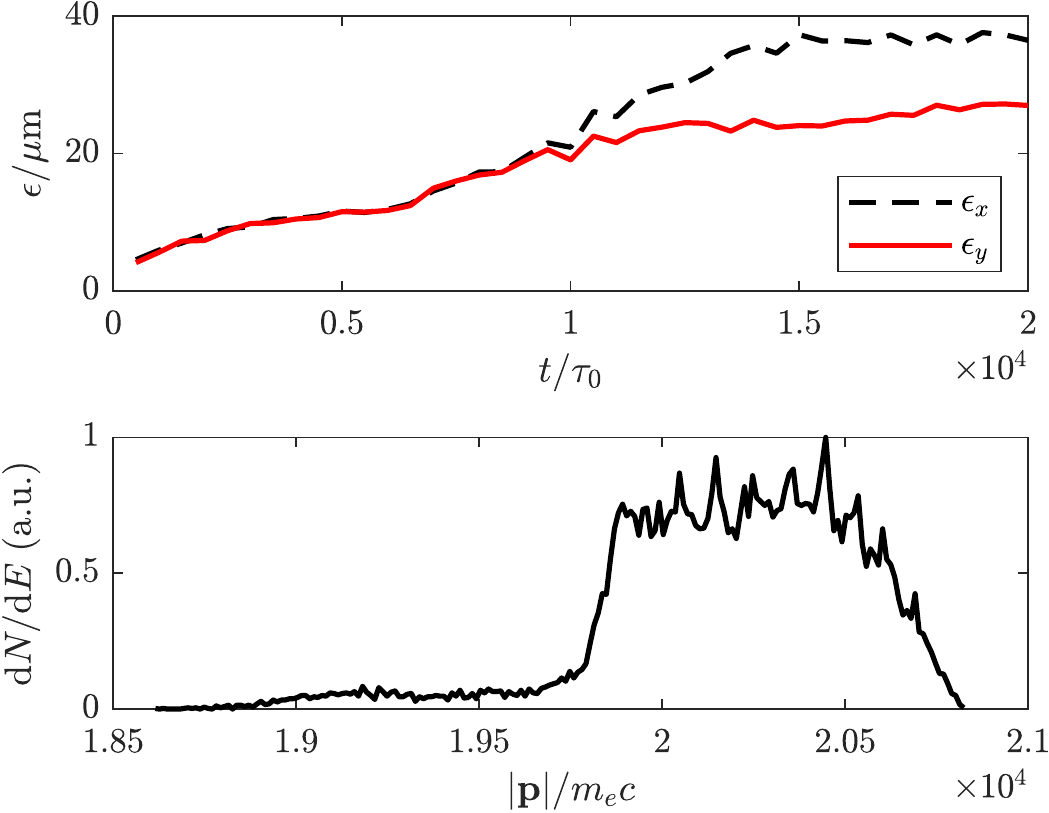}
	\caption{\label{fig:emittance}The evolution of the normalized rms emittance (top) for a positron ring with $1 n_0$. The emittance starts to increase but reaches a plateau towards the simulation end. The bottom frame shows the final energy spectrum of the accelerated positron bunch after $2 \times 10^4 T_0$.}
\end{figure}

	\begin{figure}
		\includegraphics[width=\columnwidth]{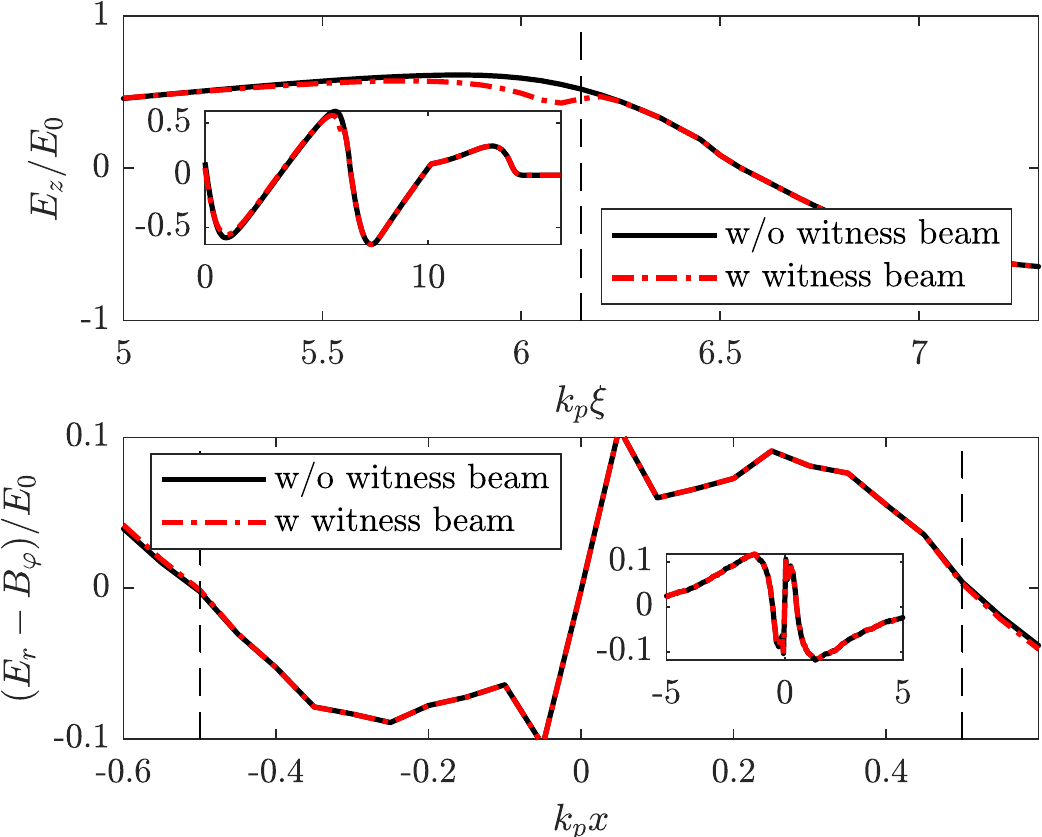}
		\caption{\label{fig:lineout}Lineout plots for the accelerating electric field (top) and the transverse force (bottom), both for the cases with witness beam (here, $10n_0$) and without. The lineouts are taken at the initial position of the position ring (further denoted by the dashed lines). The insets show the larger scale behaviour of the fields.}
	\end{figure}

	For the beam quality, we consider the normalized rms emittance $\epsilon_x = \sqrt{\langle x^2 \rangle \langle p_x^2 \rangle - \langle x p_x \rangle^2} / m_e c$ (and accordingly also in $y$-direction). Here, $x$ denotes the displacement of the particles in $x$-direction and $p_x$ the corresponding momentum component. The operator $\langle \cdot \rangle$ is the second central moment of the particle distribution. 
	As shown in Fig. \ref{fig:emittance}, the emittance increases over the course of the simulation. In the beginning, both $\epsilon_x$ and $\epsilon_y$ grow similarly. After roughly $10^4 T_0$, the difference in their slope becomes more pronounced, most likely due to the presence of plasma instabilities. Towards the end, the slope flattens and a ``saturation point'' of around $\epsilon_x  \approx 36\,\mu\text{m}$, $\epsilon_y  \approx 27\,\mu\text{m}$ is reached.
	
	In separate simulations we displace the positron ring by one grid cell in transverse and in longitudinal direction. We observe that the transverse shift only marginally affects the final emittance of the witness beam, while the setup is more sensitive for the longitudinal displacement.
	Much stronger mismatching of the positron beam size with the plasma setup leads, as mentioned earlier, to the defocusing and loss of positrons and subsequently to higher emittance. It needs to be stressed, however, that the emittance values obtained for these ring bunches are generally not comparable with different geometries like Gaussian bunches.
	
	The accelerating and focusing fields that are prominent at the position of the ring are shown in Figure \ref{fig:lineout}. We observe an accelerating gradient of $E_z \approx 38$ GV/m. The structure typical for the accelerating field $E_z$ of the wakefield can be observed. Adjusting the density and the dimensions of the positron beam leads to flattening of the accelerating field in the surrounding region.
	
	In total, we accelerate roughly $15$ pC of positrons in the case of a ring with initial density $10 n_0$ (with $n_\mathrm{driver} \approx 670$ pC). For our presented simulation, the mean positron energy is in the range of 10 GeV after acceleration over a length of 24 cm. As we observe in Fig. \ref{fig:emittance}, the final energy spread of the accelerated positrons is around $1.7\%$. Increasing the initial density of the positron ring, we are able to accelerate ca. 60 pC with the same setup. For even higher initial densities, the beam loading leads to a loss of additional charge, and the emittance and the energy spectrum deteriorate accordingly.
	
	Regarding the stability of the LAB scheme, several factors need to be discussed, the first being dephasing. Dephasing between the laser pulse and the witness bunch could severely limit the acceleration length that can be achieved. Laser refraction was -- as mentioned in the section regarding the setup -- considered for all simulations. These results are in good agreement with analytical estimates: The laser group velocity is $v_\mathrm{gr} \approx 1 - n_0 / (2 n_\crit)$, where $n_\crit = \pi m c^2 / (e^2 \lambda_L^2)$ is the critical density corresponding to the laser wavelength. For a typical 800 nm laser this would mean that over the course of the acceleration length $L_\mathrm{acc} \approx 24$ cm the laser would go back by $1.8 k_p^{-1}$ relative to the accelerated positron bunch propagating with $v_z \approx c$. The entire field structure would move backwards accordingly, and the equilibrium position for transverse focusing would become closer to the central axis. Additionally, the curvature of the ionization front increases over time. Together these two effects will lead to increased emittance and destruction of the ring structure after some time. As visible from our simulations, halving the laser wavelength to be 400 nm significantly improves on the slippage and thereby the stability of the positron ring.
	Secondly, ionization defocusing of the laser pulse can become important due to the transverse plasma density profile. According to the derivation by Gibbon \cite{Gibbon2005}, this is the case when
	\begin{align}
	    \frac{n_0}{n_\crit} > \frac{\lambda_L}{\pi z_R} \; ,
	\end{align}
	where $z_R$ is the Rayleigh length. With our parameter set this would indeed be the case, but since the LAB scheme does not rely on significantly high field strengths or very small spot sizes (for our simulations we use $w_0 = 125$ $\mu$m), ionization defocusing does not negatively impact the acceleration mechanism to a noticeable extent. We only require the laser to be strong enough for the ionization of the second column which has a large radius, namely $r_2 \gg r_1$. Further, we have shown in eq. (\ref{eq:ezeq}) that larger $r_2$ are even beneficial for this setup.
	Another possible concern for this regime could be the erosion of the electron driver head due to its propagation through unionized gas. The rate with which the ionization front is slipping back over time i.a. depends on the species of gas being ionized as well as the Lorentz factor of the bunch \cite{Li2012}. If, for longer acceleration lengths, mitigation of erosion would be necessary, the work by An \textit{et al.} \cite{An2013} presents several option how to achieve this. This i.a. includes decreasing the drive beam emittance or using lower ionization threshold gas.
	Lastly, in order to estimate the impact of the relative position between electron driver and laser pulse, we considered several simulations where the delay was increased or decreased with respect to the values presented in the manuscript above (not shown here). From this it is clear that the delay between driver and laser does not matter too much, as the fields created by the initial bubble are not particularly important for the positron dynamics.
	
	\section{Conclusion}
	In conclusion, we have shown that the laser-augmented blowout (LAB) scheme allows for the acceleration of positron rings over tens of centimeters despite the driving beam's evolution and laser refraction. The focusing fields were analytically shown to have an equilibrium line along which the positron ring can be placed. This location coincides with the maximum accelerating field for that transverse position. Emittance growth is also limited when the witness beam is properly matched to the wakefield setup.
	Future work could further investigate the stability of the scheme as well as the creation and injection of the positron ring.

    \begin{acknowledgments}
    This work has been funded in parts by DFG (project PU 213/9-1). In the part concerning the analytical model, the work was supported by the Russian Science Foundation (Grant No. 20-12-00077).
	The authors gratefully acknowledge the Gauss Centre for Supercomputing e.V.  (www.gauss-centre.eu) for funding this project (qed20) by providing computing time through the John von Neumann Institute for Computing (NIC) on the GCS Supercomputer JUWELS at J\"{u}lich Supercomputing Centre (JSC). L.R. would like to thank X.F. Shen for the helpful discussions throughout this project.
    \end{acknowledgments}

\bibliographystyle{unsrt}

\end{document}